\documentstyle[times,psfig,draft]{basi}
\def\bi{\bibitem{}}

\def\ni{\noindent}
\def\beb{\begin{thebibliography}{10}}
\def\eeb{\end{thebibliography}}
\def\bei{\begin{itemize}}
\def\eei{\end{itemize}}
\def\bef{\begin{figure}}
\def\eef{\end{figure}}
\def\ben{\begin{enumerate}}
\def\een{\end{enumerate}}
\def\beq{\begin{equation}}
\def\eeq{\end{equation}}
\def\ber{\begin{eqnarray}}
\def\eer{\end{eqnarray}}
\def\edo{\end{document}}

\begin{document}
\title[neutrinos and magnetars]{Of Neutrinos and {\it Magnetars}}
\author[Sushan Konar]%
       {Sushan Konar\thanks{e-mail:sushan@cts.iitkgp.ernet.in} \\ 
        i. Department of Physics \& Meteorology, 
        ii. Centre for Theoretical Studies, \\
        Indian Institute of Technology, Kharagpur 721302}
\maketitle
\label{firstpage}
\begin{abstract}
We discuss the nature of neutrino propagation in presence of the strong 
magnetic field of {\it magnetars}.
\end{abstract}

\begin{keywords}
neutrino--propagation: neutron star--magnetic field
\end{keywords}

\ni
Standard Model Neutrinos do not have any mass, charge or anomalous magnetic 
moment. Even massive neutrinos, beyond standard model, have negligible or 
zero magnetic moment. Evidently, neutrinos do not couple to photons in vacuum. 
However, in the presence of a thermal medium and/or an external electro-magnetic 
field $\nu-\gamma$ interaction can be mediated by $W$ and $Z$ bosons, leading 
to an effective four-Fermi vertex (Fig.1A). The charged particles, running in 
the loop, confer their electro-magnetic properties to the neutrinos
making the $\nu-\gamma$ processes sensitive to the presence of an external
magnetic field. In particular, processes like neutrino Cerenkov radiation 
($\nu \rightarrow \nu \gamma$) or plasmon decay ($\gamma \rightarrow \nu \nu$) 
would be greatly enhanced due to the presence of the strong magnetic 
field (${\cal B} \sim 10^{15}$~G) of the Magnetars and can strongly influence
the space velocity, the cooling history and even the generation of the magnetic 
field in such objects.

\bef
\begin{center}{
\psfig{file=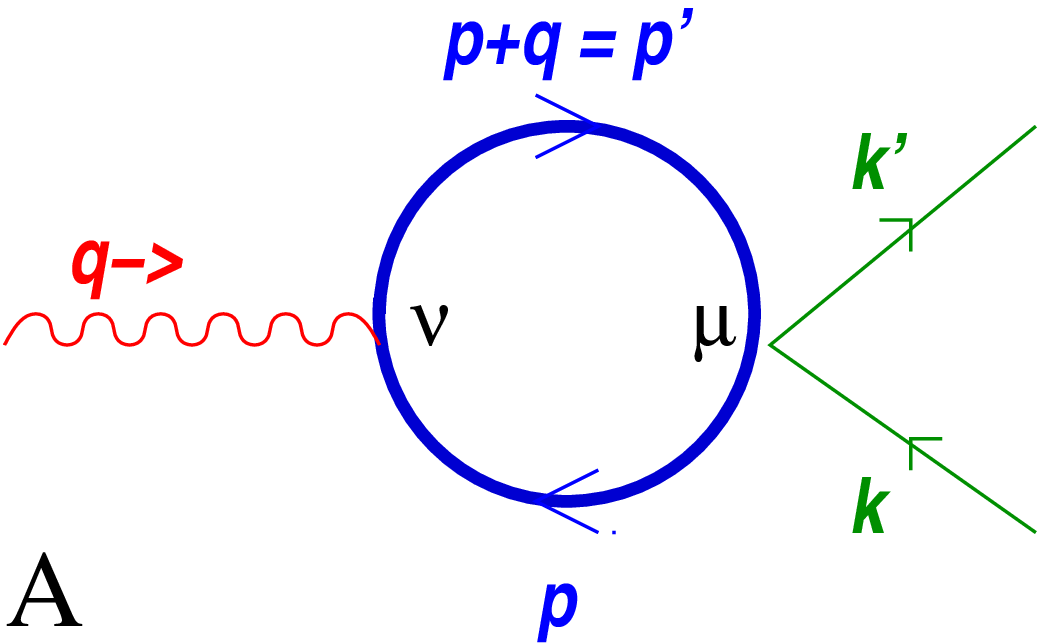,width=150pt}
\vspace{-3.25cm}                              
\hspace{5.50cm}
\psfig{file=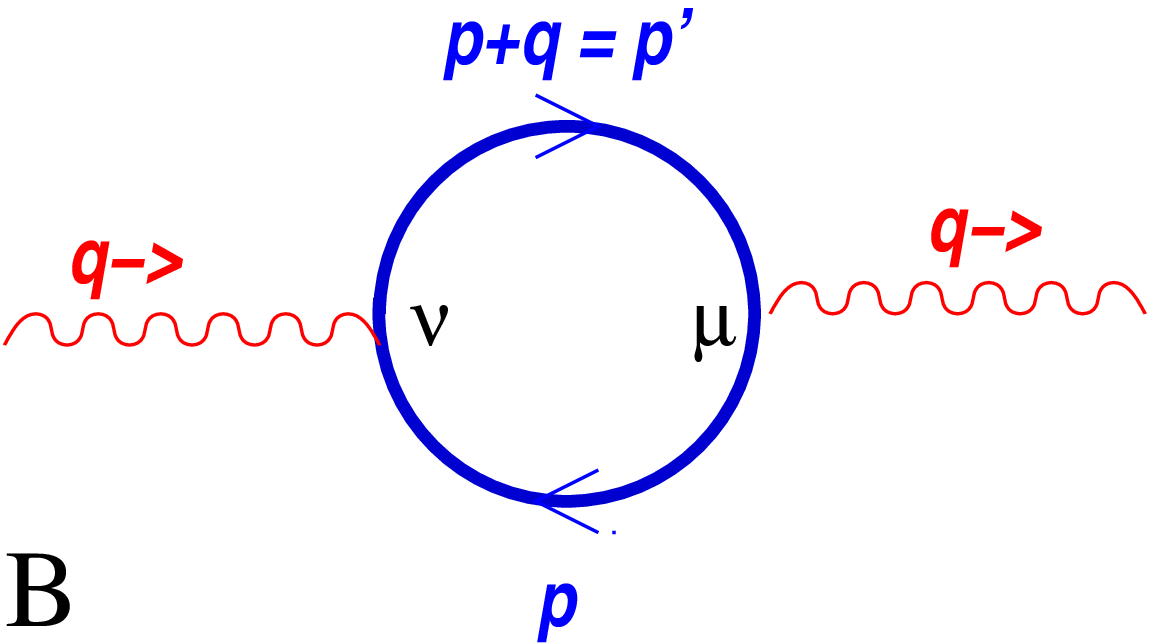,width=150pt}
}\end{center}
\caption[]{{\bf (A)} - One-loop diagram for the effective electro-magnetic 
vertex of the neutrino in the limit of infinitely heavy W and Z masses.
{\bf (B)} - One loop diagram for vacuum polarization.}
\label{fig01}
\eef

\ni 
The amplitude of a $\nu-\gamma$ process is proportional to the vertex 
function $\Gamma$ which, to leading order in the coupling constant $G_F$ and 
in the limit of $M_W, M_Z \rightarrow \infty$, is given by 
\beq
\Gamma_{\nu} = - \frac{1}{\sqrt{2}e} G_F \gamma^{\mu} (1 - \gamma_5) 
                \,(g_{\rm V} \Pi_{\mu \nu}
              - g_{\rm A} \Pi_{\mu \nu}^5)
\eeq
where, $\Pi_{\mu \nu}^5$ and $\Pi_{\mu \nu}$ are the axial-vector--vector 
coupling (Fig.1A), and the photon self-energy (Fig.1B). The dependence of
$\Gamma$ on the magnetic field and the density of the medium enter through
the electron propagators (internal lines) of the self-energy and $\nu-\gamma$
coupling diagrams of Fig.1.

\ni
Recently, Bhattacharya, Ganguly \& Konar (2002) have shown that, to the first
order in the magnetic field strength, the effective charge of the neutrinos is
\beq
e^{\nu}_{eff}({\cal B})
  \propto \Big( \frac{\cal B}{{\cal B}_e} \Big) f(m \beta) \cos \theta
     e^{\nu}_{eff}({\cal B}=0) \,, \; \; \; \;
\mbox{${\cal B}_c$ - critical field for electrons}
\eeq
where $\theta$ is the angle between ${\cal B}$ and $k$. Evidently, the 
$\nu-{\cal B}$ interaction is direction dependent which gives rise to 
asymmetries in neutrino-driven supernova explosion and may result in 
large pulsar kick velocities (Bhattacharya \& Pal, 2000). 

\ni 
The previous investigations have only been concerned with the $\nu-\gamma$
interaction to the linear order in the magnetic field strength. Such results
are pertinent for weak-field situations. However, in the context of Magnetars
we need the results valid for all orders in the magnetic field strength. In
the present work, we calculate the rate of these processes to all orders in
the magnetic field strength. Our preliminary results show that the absorptive
part of the $\nu-\gamma$ processes, to the lowest order, is enhanced by a
factor $\sim  {\cal B}^2$ (Konar \& Das, 2002). Evidently, these effects are
important only for strong fields and can be ignored otherwise.

\ni
It is understood that with the pre-existing poloidal magnetic field a strong 
toroidal field can be generated within the supernova due to the differential 
rotation near the core. The asymmetric neutrino emission would be able to provide 
the necessary torque required for this rotation and enhance the magnetic 
field to very large values (Gvozdev \& Ognev, 1999). An estimate of such
enhancement, based on our result, would be presented in a future communication
(Konar, 2003).

\beb
\bi Bhattacharya K, Ganguly A.~K. and Konar S., 2002, \newblock {\it PRD}, 
    {\bf 65}, 013007
\bi Bhattacharya K. and Pal P.~B., {\tt hep-ph/0001077}
\bi Bhattacharya K. and Pal P.~B., {\tt hep-ph/0212118}
\bi Konar S. and Das S., 2002, {\tt hep-ph/0209259}
\bi Konar S., 2003, \newblock {\it in preparation}
\bi Gvozdev A.~A., Ognev I.~S., 1999, \newblock {\it JETP Lett.}, {\bf 69}, 365
\eeb
\label{lastpage}

\end{document}